\documentclass{elsart}

\usepackage{graphicx}
\usepackage[latin1]{inputenc}
\usepackage{natbib,amssymb,graphicx}
\usepackage{pstricks} % for color definitions, to be removed !!!

\newcommand{\pT}{p_T}% {p_\perp}
\newcommand{\kT}{k_T}

\newcommand{\Pythia}{\textsc{Pythia}}
\newcommand{\Jetset}{\textsc{Jetset}}

\newcommand{\UNIT}[1]{\mbox{$\,{\rm #1}$}}

\newcommand{\GeV}{\UNIT{GeV}}

\newcommand{\fm}{\UNIT{fm}}

\newcommand{\REM}[1]{}

\begin{document}

\begin{frontmatter}

\title{Time Dependent Hadronization\\ via HERMES and EMC Data Consistency}

\author{K.~Gallmeister\corauthref{cor}}
\corauth[cor]{Corresponding author.}
\ead{Kai.Gallmeister@theo.physik.uni-giessen.de}
\author{U.~Mosel}
\address{Institut f\"ur Theoretische Physik, Universit\"at Giessen, Germany}

\begin{abstract}
  Using QCD-inspired time dependent cross sections for pre-hadrons we
  provide a combined analysis of available experimental data on hadron
  attenuation in DIS off nuclei as measured by HERMES with 12 and
  27\GeV{} and by EMC with 100 and 280\GeV{} lepton beam energies.
  We extract the complete
  four-dimensional evolution of the pre-hadrons using the
  \Jetset{}-part of \Pythia{}. We find a remarkable sensitivity of
  nuclear attenuation data to the details of the time-evolution of
  cross sections.
  Only cross sections evolving linearly in time describe the available
  data in a wide kinematical regime.
  Predictions for experimental conditions at JLAB/CLAS (5 and 12\GeV{} beam
  energies) are included.
\end{abstract}

\end{frontmatter}

%------------------------------------------------------------------------
% 1: Introduction
%------------------------------------------------------------------------
\section{Introduction}

The process of hadronization, i.e.~the question of how some partonic
state evolves into a final observed hadron wave function, is still not
understood.

For an understanding of jet interactions in hot and dense (possibly
quark-gluon) matter, investigated in ultra-relativistic heavy-ion
collisions, or even in cold nuclei, investigated e.g.\ by the EMC and the
HERMES experiments, two main features need clarification: first, the time
it takes to form a hadron needs to be known. Here we address the time
between the initial interaction until the final on-shell propagation of
the produced particle. Second, the question, how this 'unknown object'
interacts with the surrounding matter during its 'formation time' has to
be answered \cite{Kopel1}.

Since hadron production in electromagnetic interactions with the
nucleon is assumed to be simpler than e.g.~the same process in
heavy-ion reactions, deep inelastic scattering (DIS) is the process
one has to understand first, since here at least the state of the
matter in which the parton propagates is known. The essential
question then is how long it takes until the field of a knocked-out
color charge is rebuilt.

In a classical picture, the distance of two (color--)charges with
some transverse momentum $\kappa_T=\sqrt{\langle\kT{}\rangle^2}\sim
0.35\GeV$, initially set to zero, evolves linearly in time.
Therefore the cross--section is quadratic in time, $\sigma = \pi
r^2\simeq t^2$.
However, any quantum mechanical description has to respect the
uncertainty principle. This implies that the assumption of a
constant transverse momentum like the above $\kappa_T$ is not valid:
Going to very first stages with $r\to 0$ leads to $k_T\to\infty$. A
consequent consideration of consecutive transversal weakening leads
to a linear time dependence of the cross section, $\sigma\simeq t$,
as pointed out by Dokshitzer et al.~\cite{DokBuch}.

The authors of ref.~\cite{DokBuch} also stressed that
the constraints in the
above considerations leading to a linear time dependence of the
cross section may be weakened by quantum effects resulting from
non-vanishing values of the characteristic scales, as e.g.~the
squared momentum $Q^2$ in DIS, and
as a consequence, the exact time-dependence of the
cross section is expected to lie somewhere between linear and
quadratic.
Experiments should be able to tell which the better time-dependence at
a given momentum transfer is \cite{DokBuch}.

There have already been some studies employing time-dependent
cross-sections for the description of final state interactions in
ultra-relativistic heavy-ion collisions \cite{GaGrXu03,CaGaGr04}. Such
studies, however, introduce as already mentioned another unknown into the
analysis, the properties of the matter surrounding the formed hadron.
Here, in the present paper, we therefore want to use data on nuclear
attenuation of hadrons in cold nuclei, obtained by the EMC and HERMES
experiments. A similar analysis for quasi-exclusive data \cite{Q2Formula},
which is similar in spirit to ours, suffered from the lack of reliable
data at that time.

%------------------------------------------------------------------------
% 2: Model
%------------------------------------------------------------------------
\section{Time Development of Interactions}
We first stress that most reactions considered in this paper have
relatively small $\langle Q^2\rangle$ values of only $1-2\GeV^2$. The
applicability of methods of perturbative QCD thus is doubtful. This
is the common feature of the HERMES and JLAB experiments that work in
quite different energy regimes. It is thus desirable to develop a
description that works for all three energy regimes and describes the
transition from high energies to low energies correctly. Our model of
reactions on nuclei relies on a separation of processes: In step 1) the
beam lepton interacts with a nucleon. This is modeled via the \Pythia{}
\cite{Pythia} event generator assuming that this interaction with nucleons
in a nucleus is the same as that on a free proton/neutron. We do, however,
take into account nuclear effects like Fermi motion, Pauli blocking and
nuclear shadowing (for details see \cite{FaCaGaMo04a,FaCaGaMo04b,FaPHD}).

The \Pythia{} model used for step 1) has been proven to be very successful
in describing hadron multiplicities, momentum distributions etc.~in many
kinds of interactions, also at the low $Q^2$ and $\nu$ values considered
in this paper. The model combines hard physics for the initial parton
interactions and soft physics in modeling the fragmentation function. We
have therefore used \Pythia{}  as our major source of information about
the underlying process of fragmentation. The extraction of the space-time
production and formation points from \Pythia{} is described in
\cite{GaFa05}. Contrary to many analytic estimates of formation times, we
extract our information per particle per event during our Monte Carlo
simulations and do not use any averaged distributions. There is thus no
longer any freedom in choosing the relevant times in our approach. We
note that these times are model dependent, since they rely on the Lund
string fragmentation picture as implemented in \Pythia{}.
However, the simultaneous analysis of attenuation data on various
targets may help to separate effects of times from those of
pre-hadronic interactions. 
We also stress
that \Pythia{} not only contains string fragmentation, but also direct
interaction processes such as diffraction and vector meson dominance.

As elaborated in \cite{GaFa05}, in every event during the Monte
Carlo calculations and for each final particle we extract three
4D-points in \Pythia{}: First, for example, the two production points
$P1$ and $P2$ of the quark and antiquark that make up a final meson;
these are the points where the string breaks occur for each of the
two quarks. The meeting point of the quark lines starting at these
two production points is identified with the hadron formation point
(denoted by $F$). The corresponding ``formation time'' is denoted by
$t_F$.

In the following we will always identify the ``production time'' of a
particle with the ``first'' string break, i.e.~$t_P =
\min(t_{P1},t_{P2})$. We have checked that our results are frame
independent: Doing the time ordering in the laboratory frame or in
the center of momentum frame of the string has no influence.

Resulting directly from the fragmentation, any meson or baryon may consist
of 0, 1, 2 or even 3 (``leading'') partons, which build up the initial
string configurations. ``Leading'' particles have at least one parton line
directly connected with the hard interaction point and also have at least
one production time which is zero in all frames. Their properties
are, therefore, directly determined by the $Q^2$ of the primary virtual
photon. Particles with 0 leading partons, i.e.~``secondary'' or
``non-leading'' particles, all have non-vanishing production times (as
described in ref.~\cite{GaFa05} both production points are different from
the hard interaction point). We note that this picture is very similar to
that proposed by Kopeliovich et al.~\cite{Kopel1,Kopel2}.

While so far we have used only well tested in vacuo descriptions of
particle production, we now model in step 2) the in--medium changes of
the fragmentation function by introducing (pre--)hadronic interactions
between the production and the formation times.
In this step 2) all produced (pre-)hadrons are propagated through the
surrounding nuclear medium according to a semiclassical
Boltzmann--Uehling--Uhlenbeck (BUU) transport description which
allows for elastic and inelastic rescattering and side-feeding
through coupled-channel effects. For details  we refer the reader to
\cite{FaCaGaMo04a,FaCaGaMo04b,FaPHD}. For the actual numerical
treatment of final state interactions we have used the completely
rewritten, new GiBUU code \cite{GiBUU}, which is based on the theory
and methods described in \cite{FaCaGaMo04a,FaCaGaMo04b,FaPHD} and
reproduces the results presented there.

In this work we will consider four different time evolutions for the
cross sections between the production time $t_P$ and the
formation time $t_F$; after $t_F$ the hadrons interact with their full
cross section. In the first scenario we assume no time
dependence at all, i.e.~the pre-hadronic cross section is constant,
\begin{eqnarray}
  \sigma^*/\sigma = const \quad = \quad 0.5\qquad,\label{eq:scenarioC}
\end{eqnarray}
where $\sigma$ is the full hadronic cross section. Here the value
0.5 for the constant cross section ratio is chosen because it gives
a reasonable description of the HERMES data \cite{FaPHD}. The next
two scenarios are the ``quantum mechanically inspired'' and the
``naive'' assumptions of linear or quadratic increase, respectively
\begin{eqnarray}
  \sigma^*(t)/\sigma = \left(\frac{t-t_P}{t_F-t_P}\right)^n\qquad,\quad n=1,2\quad.
  \label{eq:scenarioL}
\end{eqnarray}
All three scenarios for the pre-hadronic interaction mimic to some
extent color transparency because the interaction rates are reduced
until the formation of the final hadron.

Finally, we implement the 'quantum diffusion' picture of Farrar et
al.~\cite{Q2Formula} proposed by these authors to describe the
time-development of the interactions of a point-like configuration
produced in a hard initial reaction (see also \cite{Larson}). This
picture combines the linear increase with the assumption that the
cross section for the \emph{leading} particles does not start at
zero, but at a finite value connected with $Q^2$ of the initial
interaction,
\begin{eqnarray}
\sigma^*(t)/\sigma  &=&X_0
+(1-X_0)\cdot\left(\frac{t-t_P}{t_F-t_P}\right)\qquad,\qquad X_0 =
{r_{\rm lead}}\frac{const}{Q^2}\quad, \label{eq:scenarioQ}
\end{eqnarray}
with $r_{\rm lead}$ standing for the ratios of leading partons over
the total number of partons (2 for mesons, 3 for baryons).
The baseline value $X_0$ is inspired by the coefficient $\langle n^2
k_T^2\rangle/Q^2$ in \cite{Q2Formula}.
Our scaling with $r_{\rm lead}$ guarantees
that summing over all particles in an event, on average the
prefactor becomes unity. The numerical value of the constant in the
numerator of $X_0$ is chosen to be $1\GeV^2$ for simplicity, close to
the value used in \cite{Q2Formula}.
This value is also constrained by the considered $Q^2$ range such that
the pedestal value $X_0 \le 1$ is fulfilled.
In all
four scenarios the (pre--)hadronic cross section is zero before
$t_P$ and equals the full hadronic cross section after $t_F$. The
most essential feature of color transparency -- larger hadrons (smaller
$Q^2$) get attenuated more than smaller ones -- is thus included in all
four scenarios.
Until the hadron reaches its physical groundstate the actual cross
section will oscillate around an average as pointed out by Kopeliovich et
al.~\cite{Kopel3}.

It is worthwhile to reemphasize the differences of the cross section
evolutions of leading and non--leading particles in the last model:
\emph{leading} particles start to interact with a non vanishing (i.e.~a
pedestal) cross section at the hard interaction time; they 'remember' the
$Q^2$ of the incoming photon. \emph{Non leading} particles are entirely
generated by soft string breaks, they are detached from the hard
interaction point and have no memory of the original hard interaction
process. They, therefore, start to interact at later times with zero
cross section. In both cases, the cross section increases with time. These
features reflect color transparency.

In contrast to other descriptions of the attenuation of jets in
photonuclear reactions \cite{Kopel1,Kopel2,Wang} our method
describes the whole kinematical range of final particles and is thus
not restricted to leading hadrons or very high energies only.

%------------------------------------------------------------------------
% 3: Results
%------------------------------------------------------------------------
\section{Results}

In all the following discussions we express the modification of the
spectra by the medium via the usual nuclear modification ratio
\begin{equation}
R_M^h(\nu, Q^2, z_h,\pT^2,\dots)= \frac{\
\left[N_h(\nu,Q^2,z_h,\pT^2,\dots)/N_e(\nu,Q^2)\right]_A\ }{\
  \left[N_h(\nu,Q^2,z_h,\pT^2,\dots)/N_e(\nu,Q^2)\right]_D\ }\qquad,
\end{equation}
where all the hadronic spectra on the nucleus (``A'') as also on
deuterium (``D'') are normalized to the corresponding number of
scattered electrons. As indicated, the nuclear modification ratio
can be displayed as function of many variables as e.g. $\nu$, $z_h$,
$\pT^2$  etc. Most information would be provided by multidimensional
distributions, which are, however, not yet available experimentally.

The ``photonic'' parameters of the collisions are given by $\nu$ as
photon energy and by $Q^2$ as the transferred four momentum squared.
The third parameter to fix all lepton/photon kinematics is given here
by the lepton beam energy.

The ``hadronic'' variables we focus on in this paper are $z_h$ or
$\pT^2$. Here $z_h$ stands for the ratio of the energy of the hadron
divided by the energy of the photon, while the squared transverse
momentum in respect to the photon direction is indicated by $\pT^2$.

We emphasize here our earlier findings \cite{FaPHD} that the
interpretation of the $\nu$ dependence of experimental hadron
attenuation is complicated by experimental acceptances and
integration cuts. These influence the slope of $R(\nu)$ so that a
``physical'' interpretation is possible only if these experimental
acceptance effects are taken into account in the comparison of
theory with experiment. Only a full event description such as the
one presented here can thus lead to a reliable interpretation of
experimental data. In our theory the description of the $\nu$
dependence is fixed by the description of the the $z_h$ dependence.

%\subsection{HERMES27}

\begin{figure}[tb]
  \begin{center}
    \hspace*{\fill}
    \includegraphics[height=4.5cm]{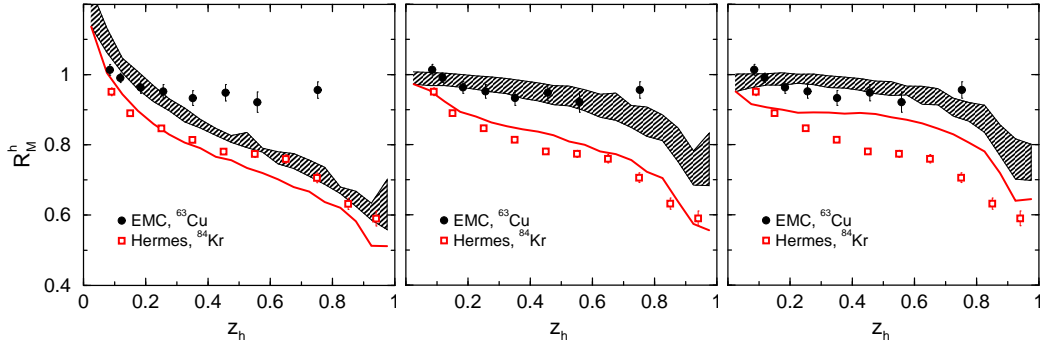}
    \hspace*{\fill}
    \caption
    {\textit{
        Nuclear modification factor for charged hadrons.
        Experimental data are for HERMES@27\GeV{}
        \cite{DataHermes} and EMC@100/280\GeV{} \cite{DataEMC}. The
        predictions for the two EMC energies are given by the
        lower and upper bounds of the shaded band.
        The cross section-evolution-scenarios in the calculations
        are: constant, linear, quadratic (from left to right).
        }}
    \label{fig:HermesEMC_1}
  \end{center}
\end{figure}

In fig.~\ref{fig:HermesEMC_1} we show for the three scenarios
according to eqs.(\ref{eq:scenarioC}),(\ref{eq:scenarioL}) (with
$n=1,2$) the results of our calculations compared to experimental
data \cite{DataHermes,DataEMC}.
Because it remains unclear to us how the (very) different lepton
energies were considered in the experimental results given in
\cite{DataEMC} we have performed the calculations for the two most
prominent energies of that experiment, i.e.~for beam energies of
$100\GeV$ and $280\GeV$. We illustrate the results of our
calculations for this experiment by a shaded band in the following
figures.

Assuming a constant cross section (leftmost panel in
fig.~\ref{fig:HermesEMC_1}), we obtain a good description of the
HERMES results, while the attenuation for the EMC experiment is much
too strong (cf.~\cite{FaPHD}).
Assuming a linear time dependence, both the HERMES and EMC
attenuation are well described\footnote{Fig.3 in \cite{DataEMC} also
contains data points leading to $R_h(z_h=0.9)=0.83$, which fits very
well into the calculated energy-band. This point is not contained in
the other figures in that paper and is, therefore, not shown in our
fig.~\ref{fig:HermesEMC_1}.}.
Going even further and assuming a quadratic time dependence
(rightmost panel in fig.~\ref{fig:HermesEMC_1}), the theoretical
attenuation is too weak both for the HERMES and for the EMC
experiment, with the discrepancy between theoretical and
experimental results being significant for the HERMES experiment.
Only the theoretical scenario with a cross section evolving linearly
in time (middle panel in fig.~\ref{fig:HermesEMC_1}) is able to
describe both data sets at the same time.

\begin{figure}[htb]
  \begin{center}
    \hspace*{\fill}
    \includegraphics[width=5.5cm,angle=-90]{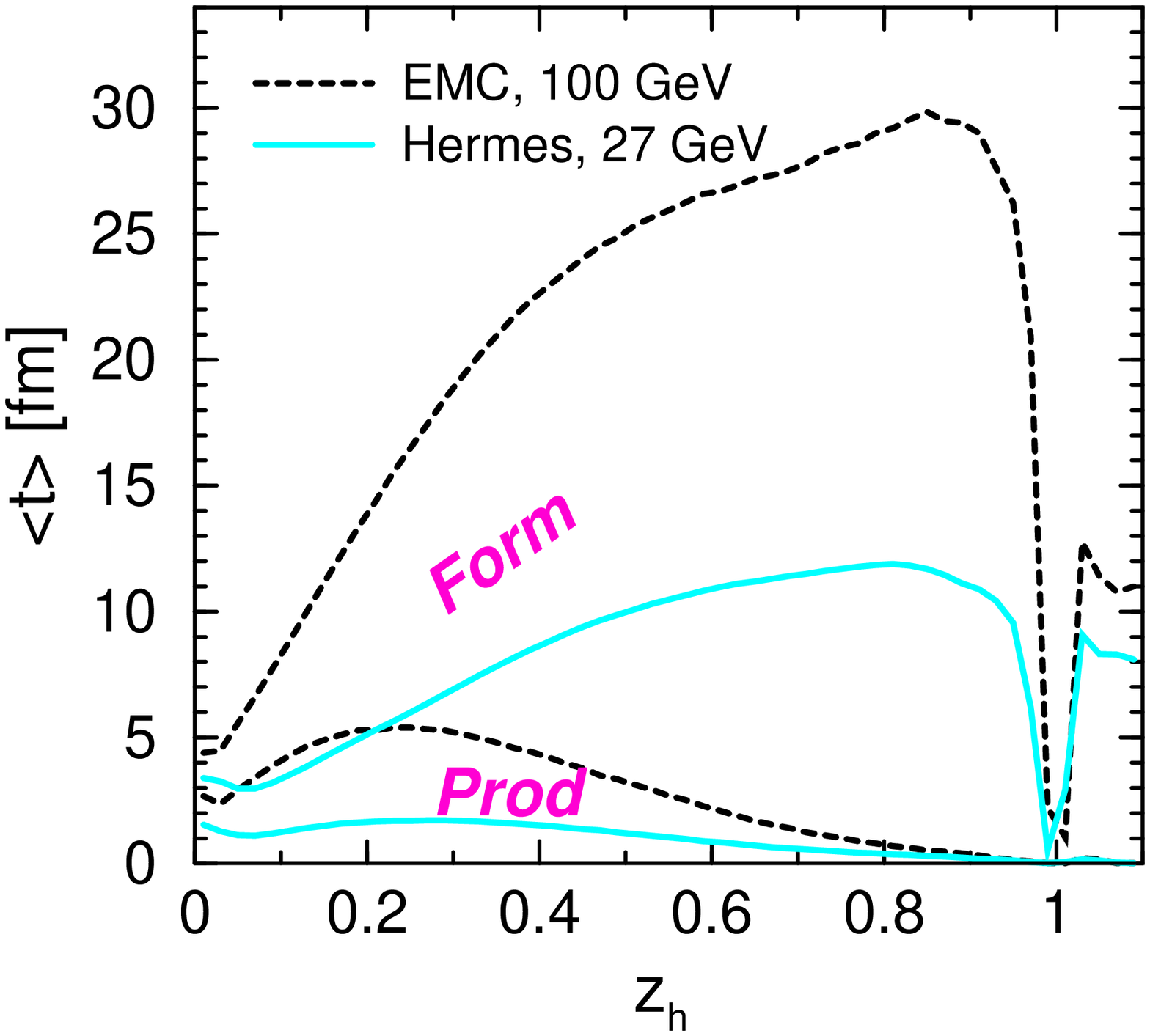}
    \hspace*{\fill}
    \includegraphics[width=5.5cm,angle=-90]{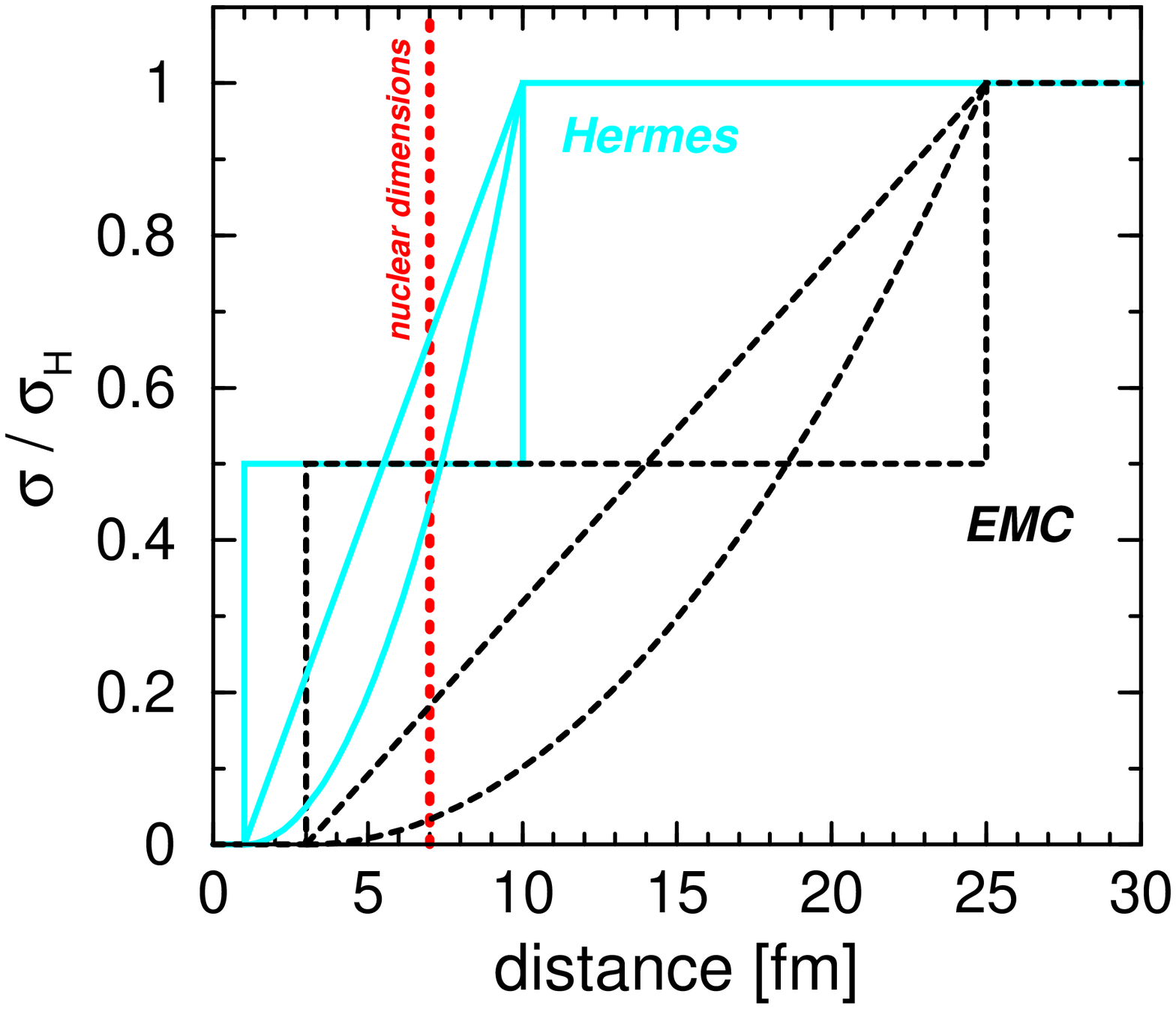}
    \hspace*{\fill}
    \caption
    {\textit{
        Left: averaged production times $\langle t_P\rangle$ and
        formation times $\langle t_F\rangle$ in the target rest frame
        for EMC@100\GeV and for HERMES@27\GeV as a
        function of $z_h$~\protect\cite{GaFa05}, averaged over leading
        ($t_P=0$) and non--leading ($t_P>0$) hadrons.
        The lowest curves give
        the production
        times whereas the two upper curves give the formation times
        for the beam energies indicated. Values of $z_h > 1$ can
        arise for baryon jets.
        Right: sketch of the evolution of the (scaled) cross section as
        function of distance from the interaction point according to scenarios
        eqs.(\ref{eq:scenarioC}) and (\ref{eq:scenarioL}) (with
        $n=1,2$): constant, linear, quadratic increase.
        The solid lines give the time--development for the HERMES
        energy regime, while the dashed lines show that for the EMC
        regime.
        }}
    \label{fig:sketchTimes}
  \end{center}
\end{figure}

In order to understand these findings, we show in
fig.~\ref{fig:sketchTimes}(a) the averaged production $\langle t_P
\rangle$ and formation times $\langle t_F\rangle$ in the target rest
frame for the two experimental setups of HERMES@27GeV and EMC@100GeV
as results of our MC calculation.
In fig.~\ref{fig:sketchTimes}(b) we sketch the different evolution
scenarios for some arbitrary chosen values of production and
formation times and compare these with a typical nuclear distance of
$\simeq 7\fm$. One sees clearly the different effects that the two
scenarios (linear and quadratic rise of cross-sections) have in the
two different kinematical regimes. For example, the quadratic
scenario leads to nearly zero interaction within the first 7\fm{}
for the EMC energy because at this higher energy the hadron has left
the nucleus before the cross section has risen to any significant
value. On the other hand, for HERMES kinematics the cross section
reaches about $0.5 \, \sigma_{H}$ in that same distance.
Figs.~\ref{fig:HermesEMC_1} and \ref{fig:sketchTimes} together show
an amazing sensitivity to the different scenarios for the
time--dependence of the cross section.
Going to lower energies than 27\GeV{} beam energy (as e.g.~with
12\GeV{} or even with 5\GeV{} lepton beam energy) results in events,
where all the hadronization happens within the nuclear distances;
at 5\GeV{} beam energy the averaged formation time is
$\simeq4\fm$ at $z_h\simeq0.8$.
In these cases we loose sensitivity to the
pre-hadronic interaction and in particular, to their time--dependence.

\begin{figure}[htb]
  \begin{center}
    \hspace*{\fill}
    \includegraphics[width=7.5cm,angle=-90]{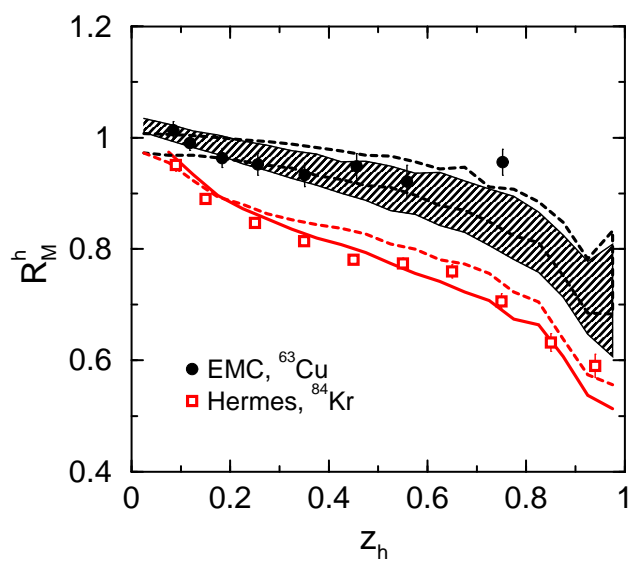}
    \hspace*{\fill}
    \caption
    {\textit{
        Nuclear modification factor for charged hadrons as in
        fig.~\ref{fig:HermesEMC_1}.
        The cross section evolution-scenario in the calculations
        is according to eq.(\ref{eq:scenarioQ}). Dashed lines repeat
        curves from fig.~\ref{fig:HermesEMC_1} (middle panel).
        }}
    \label{fig:HermesEMC_2}
  \end{center}
\end{figure}

Fig.~\ref{fig:HermesEMC_2} shows results of our calculations
employing the scenario as given by eq.~\ref{eq:scenarioQ}.
While not very pronounced, the effect of the non--vanishing,
$Q^2$--dependent initial cross section of the leading particles is
visible when comparing fig.~\ref{fig:HermesEMC_2} with the middle
panel in  fig.~\ref{fig:HermesEMC_1}; a slight improvement in the
description can be seen. The observed smallness of the $Q^2$
dependence is in line with experimental observations of both the
HERMES and the EMC experiment \cite{DataHermes,DataEMC}. This
scenario (eq.~\ref{eq:scenarioQ}) will therefore be the scenario
of our choice for the following considerations.

\begin{figure}[phtb]
  \begin{center}
    \includegraphics[width=11.6cm]{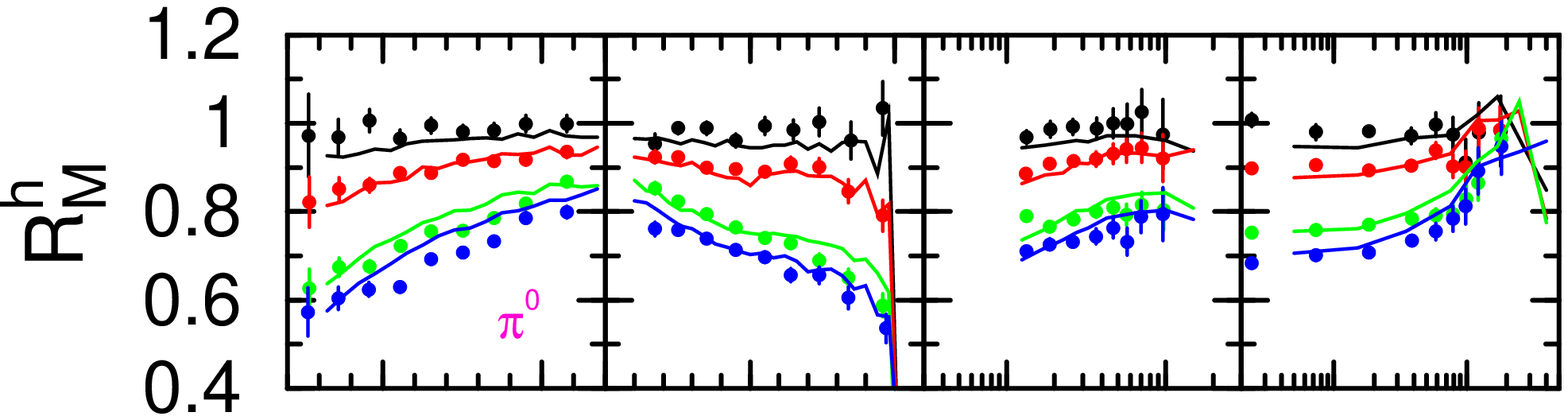}\\[-0.3cm]
    \includegraphics[width=11.6cm]{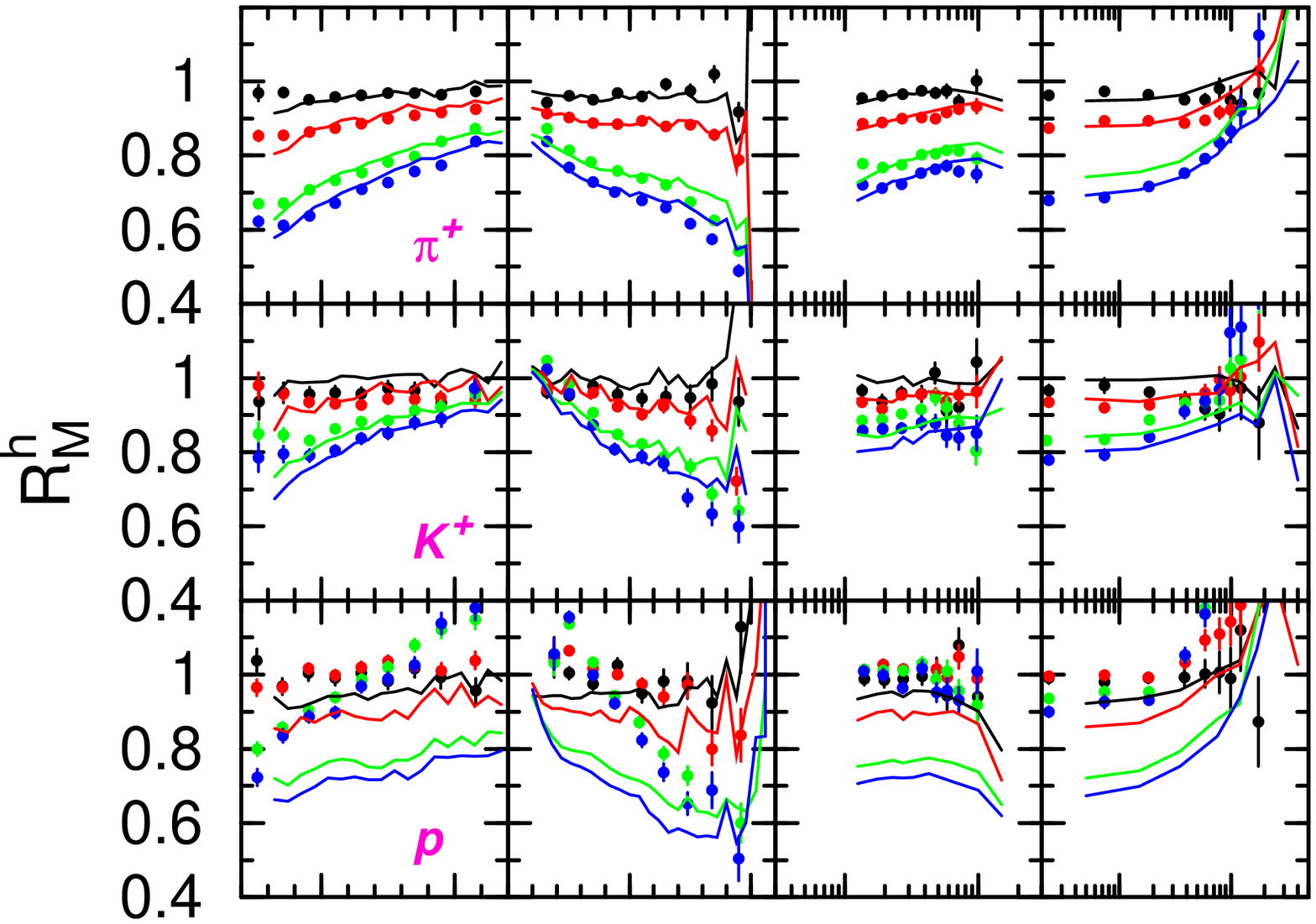}\\[-0.3cm]
    \includegraphics[width=11.6cm]{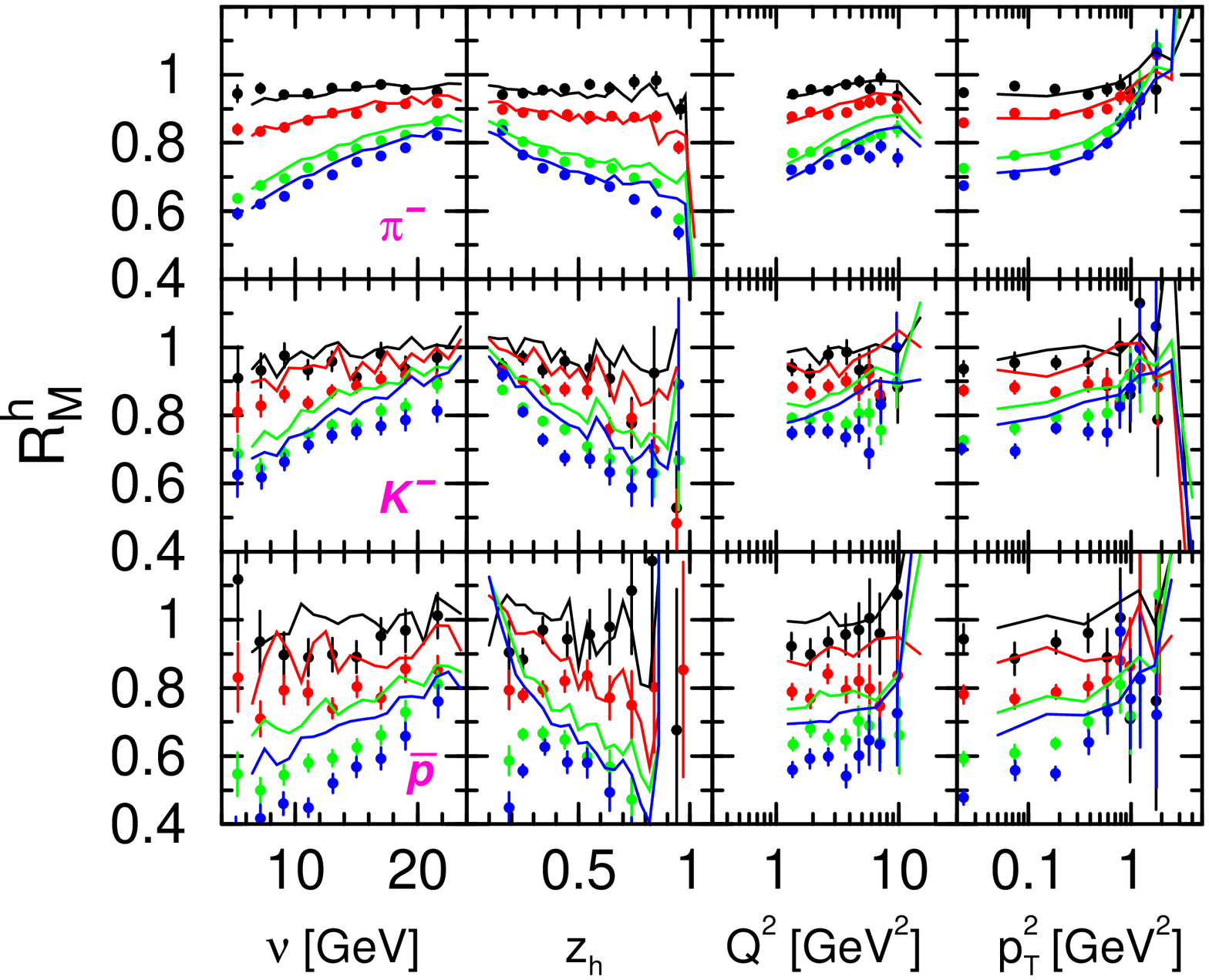}
    \caption
    {\textit{
        Nuclear modification factor for identified hadrons for
        HERMES@27GeV with $^4$He, $^{20}$Ne, $^{84}$Kr and $^{131}$Xe
        target (top to bottom).
        Points indicate experimental data \cite{DataHermes}
        while the curves represent our calculations with the
        time-dependence scenario eq.~(\ref{eq:scenarioQ}) and 
        diffractive events switched off.
        }}
    \label{fig:Hermes27Kr}
  \end{center}
\end{figure}

Fig.~\ref{fig:Hermes27Kr} shows a comparison of our calculations
with the latest experimental data of the HERMES collaboration with
27\GeV{} beam energy for identified hadrons for the four targets
$^4$He, $^{20}$Ne, $^{84}$Kr and $^{131}$Xe. 
As expected from fig.~\ref{fig:HermesEMC_2} for the total hadron
yield, the data for pions, that make up most of the produced
hadrons, are described very well by our calculations. 
In the large $z_h$ region charged pions stem mostly from decays of
diffractive rhos. Since these pions are taken out from the experimental
data, we also switch off diffractive production of rhos in the
calculations shown.
While the description of the data for the strange and anti-baryonic
sector is also quite good, one still sees the well known discrepancy
of data and calculations for protons: the regions with ``low
$z_h$''/''high $\nu$'' are clearly underestimated in our model. 
We recall that this
is not a new finding but already known from our previous work
\cite{FaPHD}. The observed discrepancy may reflect a deficiency in
our treatment of final state interactions at high proton energies
since the (strongly non-perturbative) low-$z_h$ protons arise mainly
from energy-degrading rescattering events. The discrepancy may,
however, also reflect some problem with the treatment of
experimental geometrical acceptance limitations which are contained
in the data (and simulated in the calculations).

%\subsection{HERMES12}

\begin{figure}[phtb]
  \begin{center}
    \hspace*{\fill}
    \includegraphics[width=10cm]{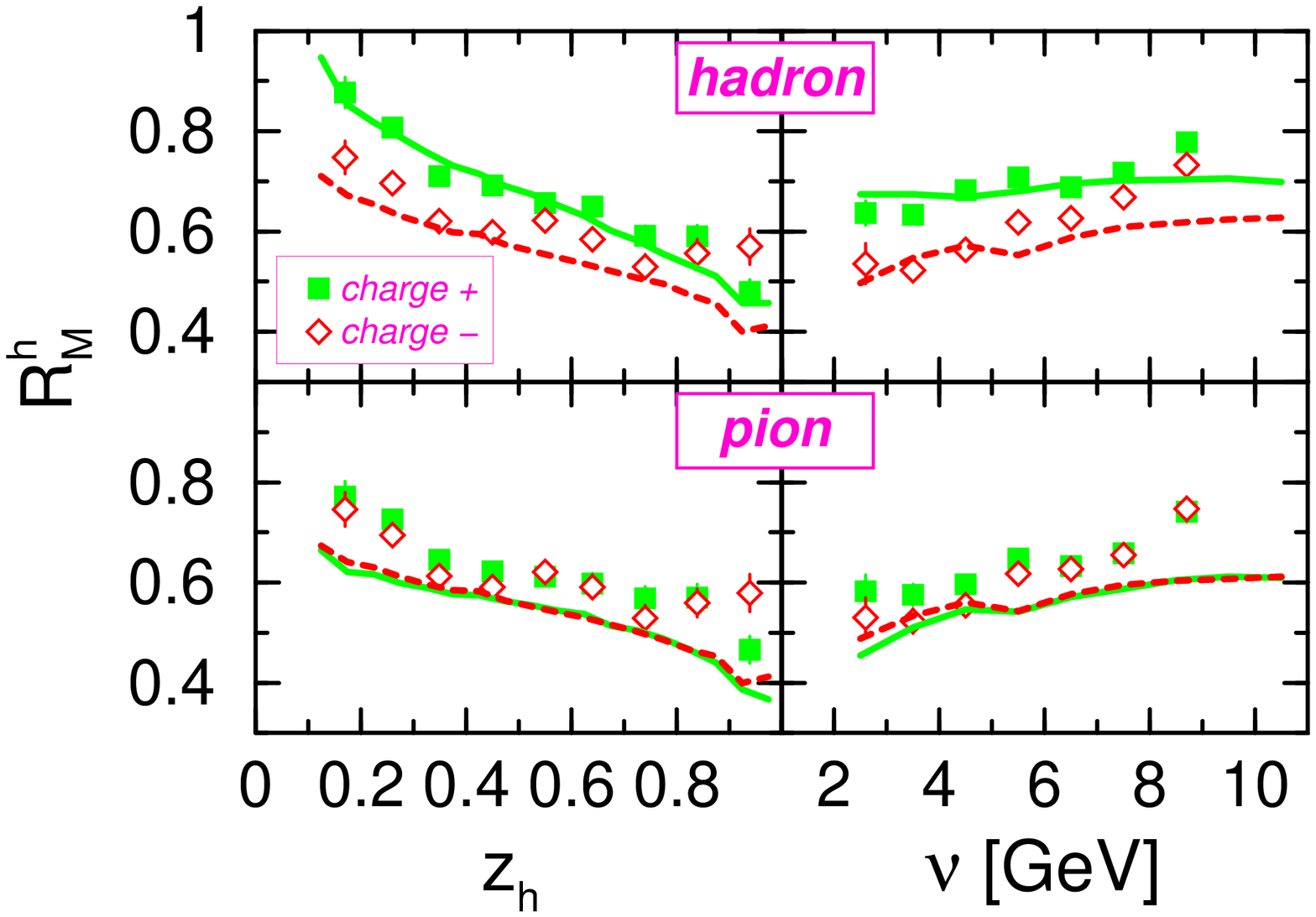}
    \hspace*{\fill}
    \caption
    {\textit{
        Nuclear modification factor of charged hadrons/pions for
        HERMES@12GeV on $^{84}$Kr target. The upper panel shows the results
        for charged hadrons, while the lower panel stands for charged
        pions only. Charge states are separated: Positive charge (full
        symbols, straight line) and negative charge (open symbols,
        dashed lines).
        }}
    \label{fig:Hermes12Kr}
  \end{center}
\end{figure}

Fig.~\ref{fig:Hermes12Kr} shows our model results compared with
experimental data of the HERMES Collaboration for the nuclear
modification factor of charged hadrons and/or pions on a $^{84}$Kr
target with the 12\GeV{} beam.
While the inclusive data for both charge states of all
charged hadrons are very well described, the attenuation for the
charged pions is somewhat overestimated.

For the $^{14}$N target all relevant lengths are smaller and all
attenuation effects are smaller: all theoretical curves are
identical to their experimental data.

%\subsection{JLAB: 12GeV, 5GeV}

Based on our successful description of the experimental data of the
HERMES collaboration for 27\GeV{} and 12\GeV{} beam energies, we now
make predictions for the meson spectra at the presently available
5\GeV{} lepton beam energy and at the future JLAB facility with
12\GeV{}. The details of the implementation of the experimental
constraints of CLAS into our MC calculations \cite{Brooks} are
described in Sect.~6.4.4 of \cite{FaPHD}. 
We note that we have made already
earlier such predictions, using constant pre-hadronic cross
sections, for the relevant JLAB energies \cite{FaPHD,Alva}.

\begin{figure}[htb]
  \begin{center}
    \hspace*{\fill}
    \includegraphics[width=10cm]{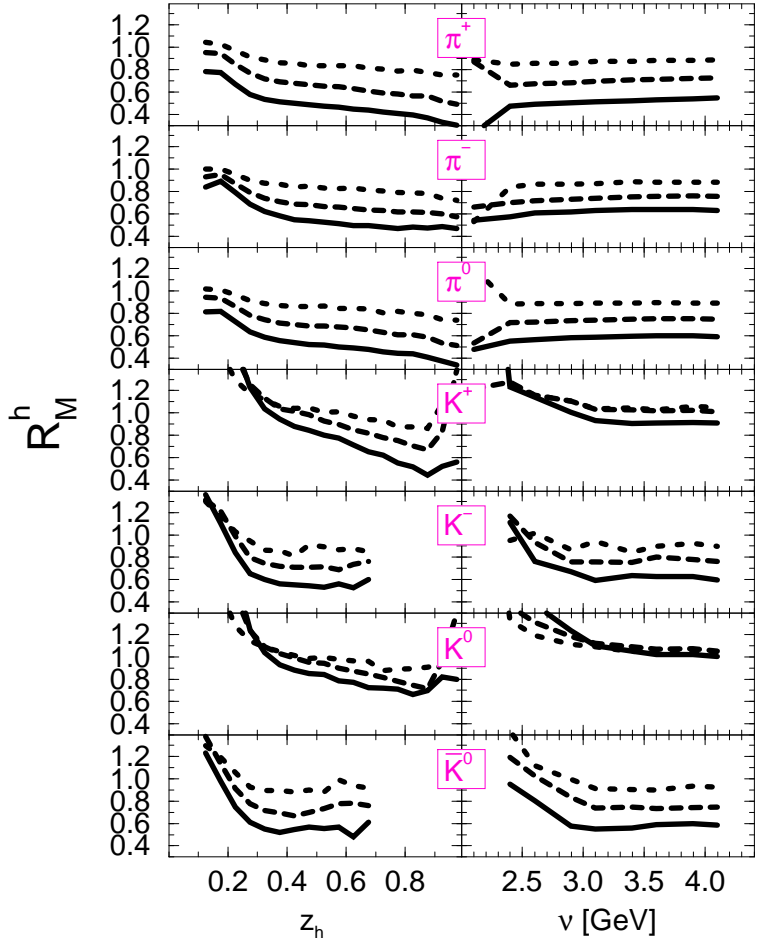}
    \hspace*{\fill}
    \caption
    {\textit{Nuclear modification factor of identified mesons
        $\pi^{\pm,0}$ and $K^{\pm,0},\bar K^0$ for
        JLAB(CLAS)@5GeV with different targets:
        $^{12}$C (dotted), $^{56}$Fe (dashed),
        $^{208}$Pb (solid lines). Experimental acceptance
        limitations are taken into account \cite{Brooks}}}
    \label{fig:JLAB5}
  \end{center}
\end{figure}

We start with a discussion of our results for a 5\GeV{} beam energy
in fig.~\ref{fig:JLAB5}.
A comparison of our results (fig.~\ref{fig:JLAB5}) with preliminary
experimental data on the $z_h$ dependence of the $\pi^+$ attenuation
for the three nuclear targets \cite{Hafidi:2006ig} is very
satisfactory, both in its magnitude and its target mass number
dependence.

\begin{figure}[htb]
  \begin{center}
    \hspace*{\fill}
    \includegraphics[width=8cm]{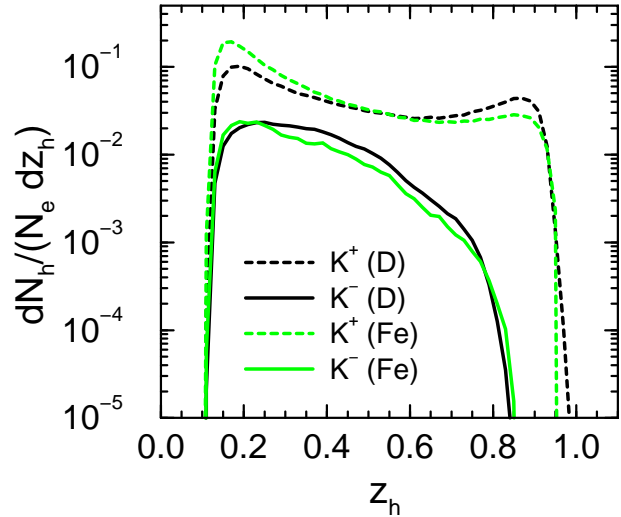}
    \hspace*{\fill}
    \caption
    {\textit{$z_h$ spectra of kaons/antikaons at JLAB@5GeV with $D$-
        and $Fe$-target.}}
    \label{fig:K-spectra}
  \end{center}
\end{figure}

Contrary to the situations at the higher beam energies, feeding
effects leading to attenuation ratios larger than unity at small
$z_h$ are more pronounced and show up to be an essential feature at
this energy. For the rarer kaons we stop showing the attenuation at
$z_h = 0.7$ because the spectra for $K^-$ drop rapidly at $z_h
\approx 0.7\dots0.8$ as shown in fig.~\ref{fig:K-spectra}.
On the contrary, the spectra for $K^+$ reach significantly farther
out.
This is a direct consequence of the fact that contrary to $K^+$ mesons
the $K^-$ mesons can only be produced in the associated strangeness
production mechanism and thus have a higher threshold than the former.
The same holds for $K^0$ and $\bar{K}^0$, respectively.

At this low energy (and corresponding momentum transfer) the
invariant masses populated in the first interaction are rather low
($\langle W\rangle=2.2\GeV$) and thus just above the resonance
region.
We have also already noted that at this low energy we have formation
times of only $\approx 4\fm$ at large $z_h$.
Therefore, the interactions
of the formed hadrons are strongly influenced by hadronic
interactions with pre-hadronic interactions playing only a minor
role, at least for the heavier targets. This shows up in
fig.~\ref{fig:JLAB5} in the different attenuation for $K^+$
and $K^-$, the latter being more strongly attenuated due to hadronic FSI.
We also recall our earlier finding \cite{FaPHD,Alva} that at this
low energy also effects of Fermi-motion are essential and have to be
taken into account. The dynamics in this energy regime is thus more
determined by 'classical' meson-nucleon dynamics than by
perturbative QCD that underlies many of the other theoretical
descriptions of the attenuation experiments
\cite{Kopel1,Kopel2,Wang}.

\begin{figure}[htb]
  \begin{center}
    \hspace*{\fill}
    \includegraphics[width=10cm]{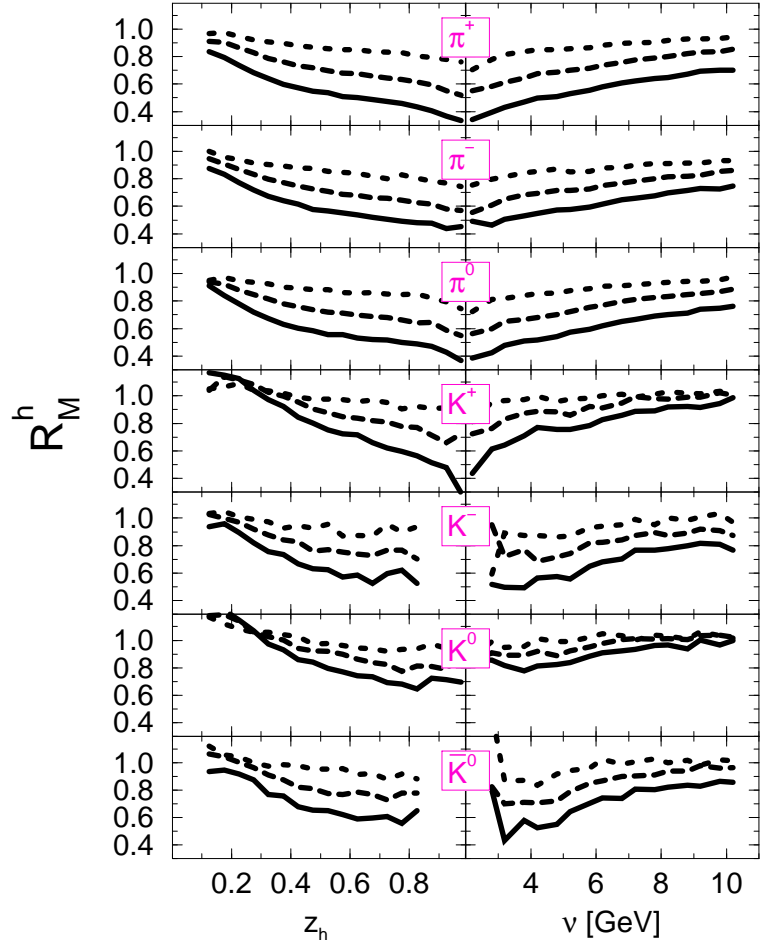}
    \hspace*{\fill}
    \caption
    {\textit{Same as fig.~\ref{fig:JLAB5}, but now for
        JLAB(CLAS)@12GeV.
         }}
    \label{fig:JLAB12}
  \end{center}
\end{figure}

Fig.~\ref{fig:JLAB12} shows the calculated results for the
multiplicity ratio of the three pion and four kaon species for the
exemplary nuclei $^{12}$C, $^{56}$Fe and $^{208}$Pb with 12\GeV{}
lepton beam energy as for the future JLAB upgrade.
For all particle species, a strong dependence of the attenuation
ratios on the size of the nucleus is obtained. It is interesting to
observe that at this higher energy the attenuation of $K^+$ and
$K^-$ is now similar at $z_h \approx 0.7$, contrary to the behavior
at 5\GeV{}. This reflects the longer formation times at this higher
energy and the corresponding predominance of pre-hadronic interactions
which affect the $K^-$ only weakly, these being non-leading hadrons.

A measurement of the $z_h$ spectra of kaons would thus give
interesting information on the production mechanism.
We note that \cite{Arleo:2003jz} have argued along similar lines.

%------------------------------------------------------------------------
% 4: Conclusions
%------------------------------------------------------------------------
\section{Conclusions}

In the present work we have calculated hadron attenuation ratios for
lepton induced reactions for the experimental conditions of the EMC
and the HERMES collaboration with lepton beam energies varying from
280/100 \GeV{} (EMC) to  27 \GeV{} (HERMES). In addition, experimental
conditions corresponding to HERMES at 12\GeV{} lepton beam energy and
JLAB (CLAS) setups with 12\GeV{} and 5\GeV{} are considered. 
We present here
a model based on (pre-)hadronic final state interactions, implemented
via coupled channel transport equations, which covers the full energy
range from 5\GeV{} up to several 100\GeV{} lepton beam energies and
reproduces all available experimental data. The model now contains
also the essential features of color transparency and should thus be
suitable to analyze future experiments searching for this phenomenon
in hadron production experiments on nuclei.

The hard interactions are described by the string-breaking mechanism
embedded in \Pythia{}. Consistent with this mechanism hadron
production and formation times are extracted from the full four
dimensional information of the \Jetset{} implementation of the Lund
string fragmentation model.

The simultaneous description of the experimental data of the HERMES
collaboration for 27\GeV{} and of the  EMC collaboration for
100/280\GeV{} lepton beam energy requires a linear increase of the
(pre--)hadronic cross section with time between the production and
formation points. The measured attenuation data are indeed somewhat
better described by pre-hadronic interactions that involve a $Q^2$
dependent pedestal value for the (pre--)hadronic cross section.
The available data show only a weak sensitivity on $Q^2$.
This may also be due to the fact that at the higher energies most of
the produced hadrons are non-leading so that they do not experience
this $Q^2$ dependence that affects only the leading hadrons at large
$z_h$.
We, therefore, expect that this sensitivity becomes stronger at lower
energies, because relatively more produced hadrons will be leading
ones.
However, at lower energies also the formation times become smaller
and, therefore, the overall influence of the pre-hadronic interaction
diminishes. In this sense the energy of 12\GeV{}, envisaged for the
JLAB upgrade, may be a good compromise between the two competing
effects. Some support for this expectation comes from our comparison
of the $K^+$ and $K^-$ attenuation. Here we have shown that at the
lower energy of 5\GeV{} the attenuation of the former is much weaker
than that of the latter because of the relatively short formation
times and the corresponding dominance of hadronic interaction. At the
higher energy of 12\GeV{}, however, the attenuation of the two becomes
similar at large $z_h$ because of the stronger influence of
pre-hadronic interactions.

It will be interesting to analyze also reactions induced by pions
($\sqrt{s}\simeq30\GeV$, Fermilab E706) or nucleons
($\sqrt{s}\simeq20\dots200\GeV$, e.g.~SPS, RHIC) to see if these
reactions with hadronic entrance channels show a different
behavior than the electromagnetic reactions analyzed in this paper
\cite{GaMo06b}.

%------------------------------------------------------------------------
% Thanks
%------------------------------------------------------------------------

\section{Acknowledgments}
This work has been supported by BMBF.
The authors thank T.~Falter for many lively discussions and his
expertise, which found its way into the actual
GiBUU code version.
We appreciate the contributions of all members of the GiBUU team,
especially O.Buss, T.Leitner, and J.Weil.
We also gratefully acknowledge support by the Frankfurt Center for
Scientific Computing.

%------------------------------------------------------------------------
%------------------------------------------------------------------------
% Literature
%------------------------------------------------------------------------


\begin{thebibliography}{99}

\bibitem{Kopel1}
B.Z. Kopeliovich et~al.,
\newblock Nucl. Phys. A740 (2004) 211.

\bibitem{DokBuch}
  Y.~Dokshitzer, V.~Khoze, A.~Mueller and S.~Troyan,
  {\it Basics of perturbative QCD}, Editions Frontiers (1991).

\bibitem{GaGrXu03}
K. Gallmeister, C. Greiner and Z. Xu,
\newblock Phys. Rev. C67 (2003) 044905.
%%CITATION = HEP-PH 0212295;%%

\bibitem{CaGaGr04}
W. Cassing, K. Gallmeister and C. Greiner,
\newblock Nucl. Phys. A735 (2004) 277.
%%CITATION = HEP-PH 0311358;%%

\bibitem{Q2Formula}
G.~R. Farrar, H.~Liu, L.~L. Frankfurt, M.~I. Strikman,
\newblock Phys. Rev. Lett. 61 (1988) 686.
%%CITATION = PRLTA,61,686;%%

\bibitem{Larson}
A. Larson, G.A. Miller and M. Strikman,
\newblock Phys. Rev. C74 (2006) 018201.
%%CITATION = NUCL-TH/0604022;%%

\bibitem{Pythia}
  T.~Sj\"ostrand et al., Comp.~Phys.~Commun.~135 (2001) 238;\\
  T.~Sj\"ostrand, L.~L\"onnblad and S.~Mrenna, LU TP 01-21, hep-ph/0108264.

\bibitem{FaCaGaMo04a}
T.~Falter, W.~Cassing, K.~Gallmeister, U.~Mosel,
\newblock Phys. Lett. B594 (2004) 61.
%%CITATION = NUCL-TH 0303011;%%

\bibitem{FaCaGaMo04b}
T.~Falter, W.~Cassing, K.~Gallmeister, U.~Mosel,
\newblock Phys. Rev. C70 (2004) 054609.
%%CITATION = NUCL-TH 0406023;%%

\bibitem{FaPHD}
  T.~Falter, Ph.D. Thesis, University of Giessen, 2004,\\
  http://theorie.physik.uni-giessen.de/documents/dissertation/falter\_phd.pdf

\bibitem{GaFa05}
K. Gallmeister and T. Falter,
\newblock Phys. Lett. B630 (2005) 40.
%%CITATION = NUCL-TH 0502015;%%

\bibitem{Kopel2}
B.Z. Kopeliovich, J. Nemchik and I. Schmidt,
\newblock Nucl. Phys. A782 (2007) 224.
%%CITATION = HEP-PH 0608044;%%

\bibitem{GiBUU}
  http://GiBUU.physik.uni-giessen.de

\bibitem{Kopel3}
B.Z. Kopeliovich and B.G.~Zakharov,
\newblock  Phys. Rev.  D44 (1991) 3466.
%%CITATION = PHRVA,D44,3466;%%

\bibitem{Wang}
E. Wang and X.N. Wang,
\newblock Phys. Rev. Lett. 89 (2002) 162301.
%%CITATION = HEP-PH 0202105;%%

\bibitem{DataHermes}
%HERMES, 12 GeV:
  HERMES, P.~van~der~Nat, Master Thesis, Amsterdam, 2002.\\
  %http://www-hermes.desy.de/notes/pub/02-LIB/PvanderNatthesis.ps.gz\\
%
  HERMES, A. Airapetian et~al.,
  \newblock Eur. Phys. J. C20 (2001) 479.\\
  %%CITATION = HEP-EX 0012049;%%
%
  HERMES, G.~Elbakyan, Proceedings of DIS2003, St.~Petersburg.\\
%
% HERMES, 27 GeV, pT:
HERMES, A. Airapetian et~al.,
\newblock Phys. Lett. B577 (2003) 37.\\
%%CITATION = HEP-EX 0307023;%%
%
HERMES, A. Airapetian et~al.,
\newblock Nucl. Phys. B780 (2007) 1.
%%CITATION = ARXIV:0704.3270;%%

\bibitem{DataEMC}
European Muon Collab, J. Ashman et~al.,
\newblock Z. Phys. C52 (1991) 1.
%%CITATION = ZEPYA,C52,1;%%

\bibitem{Brooks} W. Brooks, private communication, 2004

\bibitem{Alva}
L.~Alvarez-Ruso, T.~Falter, U.~Mosel and P.~Muehlich,
Prog. Part. Nucl. Phys. 55 (2005) 71.
%%CITATION = NUCL-TH 0412084;%%

\bibitem{Hafidi:2006ig}
CLAS, K. Hafidi,
\newblock AIP Conf. Proc. 870 (2006) 669.
%%CITATION = NUCL-EX 0609005;%%

\bibitem{Arleo:2003jz}
F. Arleo,
\newblock Eur. Phys. J. C30 (2003) 213.
%%CITATION = HEP-PH 0306235;%%

\bibitem{GaMo06b}
  K.~Gallmeister and U.~Mosel,
  in preparation.



\end{thebibliography}
\end{document}